\documentclass[conference, letterpaper]{IEEEtran}

\usepackage{fancyhdr}
\usepackage{cite}
\usepackage{amsmath,color}
\usepackage{amssymb}
\usepackage{latexsym}
\usepackage{longtable}
\usepackage{mathrsfs}
\usepackage{color}
\usepackage{balance}
\usepackage{multicol}
\usepackage{amsfonts}
\usepackage{array}
\usepackage[nomain,acronym,automake,nopostdot]{glossaries}
\definecolor{sred}{RGB}{200,21,0}
\definecolor{sblue}{RGB}{0,51,160}
\usepackage[a4 paper, top=0.75in, bottom=1in, left=0.8in, right=0.8in]{geometry}

\ifCLASSINFOpdf
\usepackage[pdftex]{graphicx}
\graphicspath{{../pdf/}{../jpeg/}}
\DeclareGraphicsExtensions{.pdf,.jpeg,.png}
\else
\usepackage[dvips]{graphicx}
\graphicspath{{../eps/}}
\DeclareGraphicsExtensions{.eps}
\fi

\newtheorem{prop}{Proposition}
\newtheorem{defn}{Definition}

\newcommand{\matc}[1]{\mbox{\boldmath $\mathcal{#1}$}}

\newcommand{\figref}[1]{Fig. \ref{#1}}

\usepackage{bbm}
\usepackage{setspace}
\usepackage[ruled]{algorithm2e}
\usepackage{amsmath}
\makeglossaries
\newacronym{ils}{ILS}{integer least-squares}
\newacronym{snr}{SNR}{signal-to-noise ratios}
\newacronym{mlsd}{MLSD}{maximum-likelihood sequence detection}
\newacronym{map}{MAP}{maximum a posteriori}
\newacronym{lmmse}{LMMSE}{linear minimum mean-square error}
\newacronym{ut}{UT}{user terminal}
\newacronym{uts}{UTs}{user terminals}
\newacronym{bs}{BS}{base station}
\newacronym{ann}{ANN}{artificial neural network}
\newacronym{dnn}{DNN}{deep neural network}
\newacronym{mimo}{MIMO}{multiple-input multiple-output}
\newacronym{mumimo}{MU-MIMO}{Multiuser multiple-input multiple-output}
\newacronym{mud}{MUD}{multiuser detection}
\newacronym{amp}{AMP}{approximate message passing}
\newacronym{zf}{ZF}{zero forcing}
\newacronym{csir}{CSIR}{channel state information at the receiver}
\newacronym{bp}{BP}{back propagation}
\newacronym{awgn}{AWGN}{additive white Gaussian noise}
\newacronym{ai}{AI}{artificial intelligence}
\newacronym{csi}{CSI}{channel state information}
\newacronym{ber}{BER}{bit error rate}
\newacronym{sgd}{SGD}{stochastic gradient descent}
\newacronym{ml}{ML}{machine learning}
\newacronym{rls}{RLS}{recursive least-square}
\newacronym{iui}{IUI}{inter-user interference}
\newacronym{sjcm}{SJCM}{systematic joint coding and modulation}
\newacronym{nsjcm}{NSJCM}{non-systematic joint coding and modulation}
\newacronym{dof}{DoF}{degress of freedom}
\newacronym{ls}{LS}{least squares}
\begin{document}
	
	\title{\huge On Deep Learning Solutions for Joint Transmitter and Noncoherent Receiver Design in MU-MIMO Systems}

	\author{\IEEEauthorblockN{Songyan Xue, Yi Ma, Na Yi, and Rahim Tafazolli}
		\IEEEauthorblockA{Institute for Communication Systems (ICS),
			University of Surrey, Guildford, England, GU2 7XH\\
			E-mail: (songyan.xue, y.ma, n.yi, r.tafazolli)@surrey.ac.uk}}
	\markboth{}%
	{}
	\maketitle
	
	\begin{abstract}
		This paper aims to handle the joint transmitter and noncoherent receiver design for multiuser multiple-input multiple-output (MU-MIMO) systems through deep learning. Given the deep neural network (DNN) based noncoherent receiver, the novelty of this work mainly lies in the multiuser waveform design at the transmitter side. According to the signal format, the proposed deep learning solutions can be divided into two groups. One group is called pilot-aided waveform, where the information-bearing symbols are time-multiplexed with the pilot symbols. The other is called learning-based waveform, where the multiuser waveform is partially or even completely designed by deep learning algorithms. Specifically, if the information-bearing symbols are directly embedded in the waveform, it is called systematic waveform. Otherwise, it is called non-systematic waveform, where no artificial design is involved. Simulation results show that the pilot-aided waveform design outperforms the conventional zero forcing receiver with least squares (LS) channel estimation on small-size MU-MIMO systems. By exploiting the time-domain degrees of freedom (DoF), the learning-based waveform design further improves the detection performance by at least 5 dB at high signal-to-noise ratio (SNR) range. Moreover, it is found that the traditional weight initialization method might cause a training imbalance among different users in the learning-based waveform design. To tackle this issue, a novel weight initialization method is proposed which provides a balanced convergence performance with no complexity penalty.
	\end{abstract}
	
	\IEEEpeerreviewmaketitle
	
	\section{Introduction}\label{sec1}
	\gls{mumimo} systems have been extensively studied in the past few decades for their advantages on increasing system capacity, enhancing spectrum efficiency and improving link reliability \cite{7244171}. A fundamental problem for \gls{mumimo} system is to estimate the transmitted signal relying on the knowledge of the received signal and the channel. To elaborate a little further, if the instantaneous \gls{csi} is available at the receiver, the detection of the transmitted signal relies on coherent detection; and there exists a wide range of solutions in the traditional communication domain \cite{8804165}. Things become difficult when the \gls{csi} is avoided, i.e., \gls{mumimo} signal detection belongs to the family of noncoherent detection \cite{6780655}. In this case, one of the most commonly used methods is called differential encoding, which imposes correlation among the transmitted symbols \cite{DBLP:journals/corr/AlsifianyIC17}. And the receiver needs to use sequence-level differential detection which leads to a higher computational complexity and a degraded power efficiency compares with the symbol-level coherent detection algorithms. Therefore, noncoherent detection is normally considered to suffer an inherent performance loss unless the block size is sufficiently large \cite{7244171}.  
	
	Recent advances towards noncoherent \gls{mumimo} signal detection lies in the use of deep learning technique. The basic idea is to model the entire \gls{mumimo} system as a DNN, then end-to-end optimization can be implemented by using deep learning algorithms. A relatively comprehensive state-of-the-art review can be found in \cite{8666641}. Notably, a joint modulation and decoding optimization algorithm is proposed in \cite{2019arXiv190303711W}, where the residual multilayer perceptron (ResMLP) is employed for modulation constellation design. In \cite{9013199}, an autoencoder-based structure is proposed to generate noncoherent space-time codes. In \cite{8385498}, compressed sensing (CS) based algorithm is proposed to tackle the noncoherent detection problem particularly in massive machine-type communications (mMTC). Apart from the existing literature, our preliminary work in \cite{8437142} provides an end-to-end optimization solution for the \gls{mumimo} systems.

	Despite their advantages, current deep learning-based joint transmitter and noncoherent receiver optimization approaches are still challenged by the signal processing scalability with respect to the size of \gls{mumimo} networks. It has been shown that most of the existing algorithms can only perform well when the size of the \gls{mumimo} system is relatively small (e.g. $2 \times 4$ or even smaller). When the spatial-domain user load increases, their detection performance decreased rapidly due to the increasing \gls{iui}.
	
	Motivated by this observation, we fundamentally rethink the design of deep learning-based \gls{mumimo} transmitter and noncoherent receiver. By exploiting the time-domain \gls{dof}, more robust solutions have been proposed towards this problem. Main contributions of this paper include:
	\begin{itemize}
		\item Two groups of deep learning based joint transmitter and noncoherent receiver optimization approaches for \gls{mumimo} systems. One group is called pilot-aided waveform, which is motivated by the traditional pilot-based waveform design. The key novelty of this approach mainly lies in the receiver side, since the received pilot symbols are utilized for DNN-based sequence-level signal detection rather than channel estimation. The other group is called learning-based waveform, where the multiuser waveform is partially or even completely designed by using neural networks. And the novelty of this approach is mainly focused on the multiuser waveform design at the transmitter side. It is shown that the pilot-aided waveform design outperforms the conventional \gls{ls} channel estimation solution on small-size \gls{mumimo} systems. Meanwhile, the learning-based waveform design largely improves the traditional channel estimation based solutions in various communication scenarios. Beside, all proposed approaches bypass channel inversion or factorization which are needed for most of the noncoherent detection approaches. Therefore, the required computational complexities are much lower.
		\item The development of a novel weight initialization method for the proposed learning-based waveform design, termed symmetrical-interval initialization. It is found that the traditional weight initialization method can result in a training imbalance among different users. To tackle this issue, the proposed approach restricts the intervals for weight initialization and yields a much better convergence performance in \gls{dnn} training phase.
	\end{itemize}

	\section{System model and Problem Statement}\label{sec2}
	\subsection{\gls{mumimo} Uplink Model}
	Consider \gls{mumimo} uplink communications, where $M$ user terminals (UTs) simultaneously communicate to an uplink access point (AP) with $N$ receive antennas $(N\geq M)$. The \gls{mumimo} channel is assumed to be block-fading, i.e., the channel remains unchanged within the coherence time $T$, and each UT employs a single transmit-antenna \footnote{We assume each user having a single antenna to focus our presentation on the key ideas. An extension to multiple antennas is trivial.} to send a temporal sequence. For each transmission block of duration $T$, the received signal block is represented by
	\begin{equation}\label{eq01}
	\mathbf{Y} = \mathbf{X}\mathbf{H} + \mathbf{V}
	\end{equation}
	where $\mathbf{Y} \in \mathbb{C}^{T\times N}$ stands for the received signal block over $T$ time slots, $\mathbf{X} \in \mathbb{C}^{T\times M}$ for the transmitted signal block, $\mathbf{H} \in \mathbb{C}^{M\times N}$ for the \gls{mumimo} channel matrix, and $\mathbf{V} \in \mathbb{C}^{T\times N}$ for the \gls{awgn} with each element independently drawn from $CN(0,\sigma^2\mathbf{I})$. Moreover, $\mathbf{I}$ is the identity matrix. Let $\mathbf{x}_m\in \mathbb{C}^{T\times 1}$ be the $m^\mathrm{th}$ column of $\mathbf{X}$. The average transmission power of each UT is regularized by
	\begin{equation}\label{eq02}
	\frac{1}{T}\mathbb{E}\left [ \mathbf{x}_m^H \mathbf{x}_m^{} \right ] \leq \delta_m P,~_{m\in\left \{ 1,2,\dots,M \right \}}
	\end{equation}
	subject to $\delta_m \geq 0$ and $\sum_{m=1}^{M} \delta_m= 1$, where $P$ is the total power budget, $\mathbb{E}(\cdot)$ is the expectation, and $[\cdot]^H$ is the matrix Hermitian.
	
	\subsection{Ambiguities in Noncoherent Detection}
	Suppose: A1) the channel matrix $\mathbf{H}$ is unknown at the receiver side, and A2) elements in $\mathbf{x}_m$ can be mutually correlated. The receiver aims to reconstruct the transmitted signal block $\mathbf{X}$ from $\mathbf{Y}$ through noncoherent sequence detection. The \gls{map} estimates of $\mathbf{X}$ is given by
	\begin{equation}\label{eq03}
	\hat{\mathbf{X}} = \underset{\mathbf{X}}{\arg \max} ~p_{\mathbf{X|Y}}(\mathbf{X|Y})
	\end{equation}
	subject to $\mathbf{X}$ drawn from a finite-alphabet set $\matc{\mathcal{A}}=\left \{\mathbf{\Theta}_1,\dots,\mathbf{\Theta}_J\right \}$, where $\mathbf{\Theta}_j \in \mathbb{C}^{T \times M}$ and $J$ the size of $\matc{\mathcal{A}}$. The $m^\mathrm{th}$ column of $\mathbf{\Theta}_j$ is the $m^\mathrm{th}$ UT's codeword, denoted by $\mathbf{c}_m$, which is independently drawn from their specific codebooks $\mathbf{C}_m$. Assuming each UT's codebook having $L$ codewords, then we have $J=L^M$.
	
	Theoretically, there are two factors that might cause a detection error in \eqref{eq03}. One is the white Gaussian noise $\mathbf{V}$, and the other is the \gls{mumimo} channel $\mathbf{H}$ (i.e. channel ambiguity). The noise effect has been well studied in \cite{8054694}, our focus is mainly on the channel ambiguity which should be well addressed in the proposed approaches.
	
	\begin{defn}[Channel Ambiguity]\label{defn1}
		It is called channel ambiguity that the transmitted signal block $\mathbf{X}$ is not uniquely determined by the received signal block $\mathbf{Y}$ even in the noiseless case. This is one of the dominating factors that limits the application of noncoherent signal detection in \gls{mumimo} systems.
	\end{defn}
	
	A natural solution towards this problem is to send a pilot sequence to estimate the fading coefficients \cite{1193803}, and then use the estimated channel to communicate. Although the channel estimation solution is outside the scope of noncoherent detection, it is closely related to the proposed pilot-aided waveform approach due to the same waveform format. Therefore, we provide a brief analysis to improve the readability of this paper.
	
	Consider the short coherence time case, where the coherent interval $T<K+N$ with $K=\min\{M,N\}$. The training phase lasts $T_\tau$ time samples and the time duration for data transmission equals to $T_d = T-T_\tau$. In order to estimate the $N$-by-$M$ channel coefficients, we will need at least $(N)\times (M)$ measurements at the receiver \cite{978730}. Therefore, the optimum length of the training interval is $T_\tau = M$ and the total \gls{dof} for communication is at most
	\begin{equation}\label{eq04}
	D_\text{pilot} = M(1-\frac{M}{T})
	\end{equation}
	and the result capacity lower bound with optimum power allocation at high \gls{snr} is given by
	\begin{align}\label{eq05}
	C_\text{pilot} \geq \big(1&-\frac{M}{T}\big)\mathbb{E} \Bigg[\log\det\bigg(\mathbf{I}_M \nonumber\\ &+\frac{1}{\Big(\sqrt{1-\frac{M}{T}}+\sqrt{\frac{M}{T}}\Big)^2} \cdot \frac{\hat{\mathbf{H}}\hat{\mathbf{H}}^H}{M} \cdot\rho\bigg)\Bigg]
	\end{align}
	where $\rho$ is the expected received \gls{snr} at each receive antenna, and $\hat{\mathbf{H}}$ the normalized channel estimate \cite{1193803}. Intuitively, the training based algorithm suffers a fractional \gls{dof} gap of $(M)/(T)$ from the ideal \gls{dof}. In Section III, we will demonstrate that the proposed learning-based approaches have the ability to mitigate the \gls{dof} loss.
	
	\section{Deep Learning based Joint Transmitter and Noncoherent Receiver Design} 
	\subsection{Pilot-aided Waveform Design}
	\begin{figure*}[!t]
		\centering
		\includegraphics[width=6.6in]{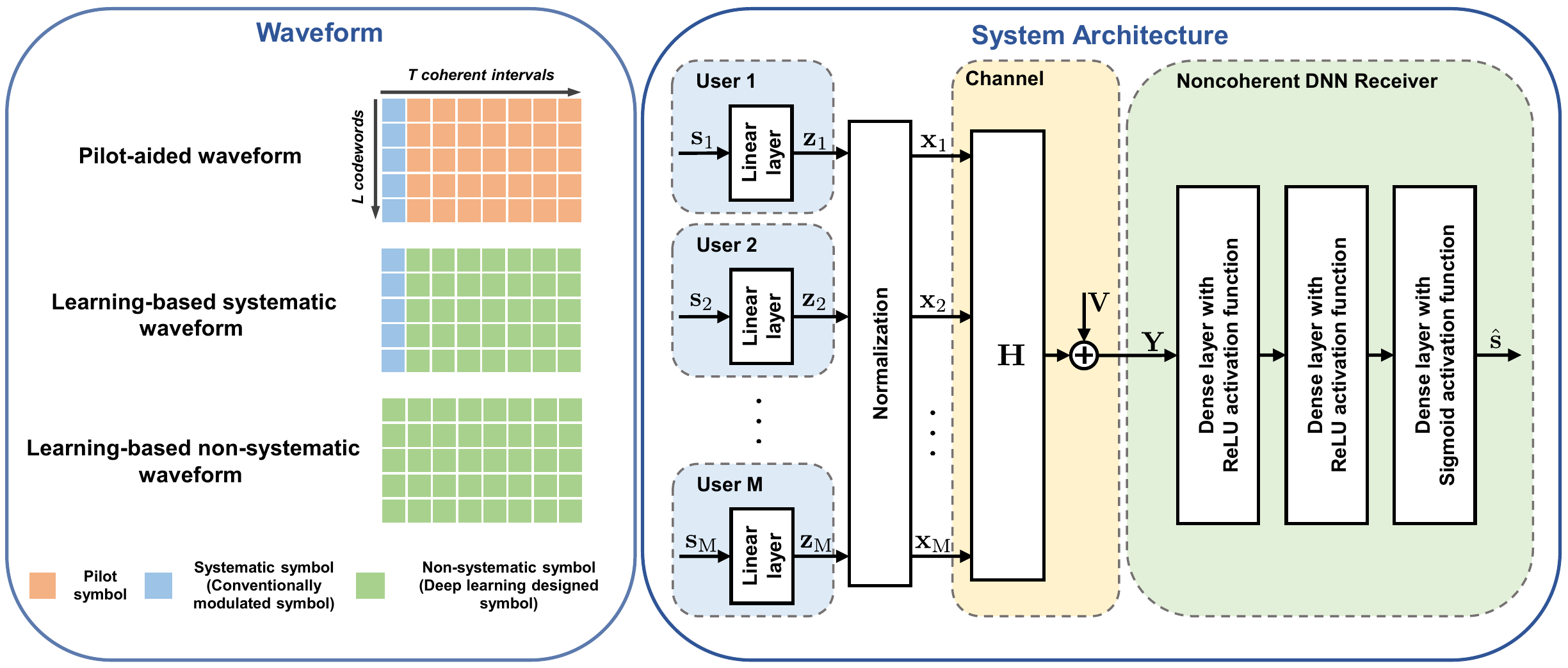}
		\caption{The proposed system architecture and waveforms for \gls{mumimo} joint transmitter and noncoherent receiver optimization.}
		\label{figure1}
	\end{figure*}
	
	\begin{defn}\label{defn2}
		It is called pilot-aided waveform design when the multiuser waveform is formed by the traditional \gls{mumimo} pilot symbols and the information-bearing symbols; as shown in \figref{figure1}.
	\end{defn} 
	
	We start from the introduction of the basic signal model, the two-part training and data process is equivalent to partitioning the matrix in \eqref{eq01} as
	\begin{equation}\label{eq06}
	\mathbf{X} = \binom{\mathbf{X}_\tau}{\mathbf{X}_d},~\mathbf{Y} = \binom{\mathbf{Y}_\tau}{\mathbf{Y}_d}
	\end{equation}
	where $\mathbf{X}_\tau$ and $\mathbf{X}_d$ stand for the matrices of training and data samples, respectively; and $\mathbf{Y}_\tau$ and $\mathbf{Y}_d$ for the corresponding received signal blocks. As aforementioned, $\mathbf{X}_d$ might not be uniquely determined by $\mathbf{Y}_d$ due to channel ambiguity. Later, we will introduce how training matrices $\mathbf{X}_\tau$ and $\mathbf{Y}_\tau$ can be used to mitigate the uncertainty.
	
	For pilot-aided waveform design, the transmitter side does not contain any neural networks. The linear layers are a group of predefined codebooks $\mathbf{W}_m^{(p)}\in \mathbb{C}^{L \times T},~_{1\leq m\leq M}$, with each consist of both pilot component and data component as described in \eqref{eq06}. The input to the linear layer is a one-hot vector, denoted by $\mathbf{s}_m,~_{1\leq m\leq M}$, with the size of $(L) \times (1)$, and $\mathbf{s}_m$ follows the uniform distribution. The output is $\mathbf{z}_m=\mathbf{W}_m^{(p)T}\mathbf{s}_m,~_{1\leq m\leq M}$, which is one of the codeword in the predefined codebook. Intuitively, $\mathbf{s}_m$ serves as a waveform selector. By combining the codewords from all $M$ users, we are able to form the signal block $\mathbf{X}$. At the receiver side, \gls{dnn} plays a central role for multiuser noncoherent signal detection. The input to the \gls{dnn} receiver is obtained by reshaping the signal matrix $\mathbf{Y}$ into a column vector $\mathbf{y}$. It is worth noting that the communication signals are normally modeled as complex-valued symbols, but most of the deep learning algorithms are based on real-valued operations. To facilitate the learning and communication procedure, it is usual practice to convect complex signals to their real signal equivalent version by concatenating their real and imaginary parts \footnote{For the sake of mathematical notation simplicity, we will not use doubled size for the rest of this paper.} (see \cite{8761999,2020arXiv200210847X,8646357,8642915,8761937,8580824,2020arXiv200400404X}) which can be described as
	\begin{equation}\label{eq07}
	\mathbf{y}_\text{real} = \begin{bmatrix}
	~\Re(\mathbf{y})~\\ 
	~\Im(\mathbf{y})~
	\end{bmatrix}
	\end{equation}
	The noncoherent DNN receiver consists of three layers: two dense layers with ReLU activation function followed by a dense layer with Sigmoid activation function. The output of the \gls{dnn} receiver $\hat{\mathbf{s}}$ is an estimate of the original information-bearing bits. Such an estimate can be trained by using supervised learning algorithms with the objective of minimizing the difference between the network estimate $\hat{\mathbf{s}}$ and the ground-truth training label $\mathbf{s}$, where the latter can be obtained by using the codebook combination approaches (see [19, Section II-B\&C] for more detailed descriptions).
		
	\begin{prop}\label{prop1}
		Given the received signal $\mathbf{Y}$ and the supervisory training target $\mathbf{s}$, deep learning will establish the link between $\mathbf{Y}$ and $\mathbf{X}_d$ according to the \gls{map} probability $p(\mathbf{s}|\mathbf{Y})=p(\mathbf{X}_d|\mathbf{Y})$.
	\end{prop}
	
	For a fixed channel matrix $\mathbf{H}$, we can easily obtain the estimate of $\mathbf{X}_d$ according to the \gls{map} probability $p(\mathbf{X}_d|\mathbf{Y}_d)$ even without the pilot component. Problem becomes difficult when $\mathbf{H}$ is randomly varying. In this case, $\mathbf{Y}_\tau$ can be employed to mitigate the channel uncertainty in noncoherent signal detection. From mutual information point of view we have 
	\begin{align}\label{eq08}
	I(\mathbf{Y}_\tau,\mathbf{Y}_d;\mathbf{X}_d) &= I(\mathbf{Y}_d;\mathbf{X}_d| \mathbf{Y}_\tau) + I(\mathbf{Y}_\tau;\mathbf{X}_d) \nonumber \\
	&=I(\mathbf{Y}_d;\mathbf{X}_d| \mathbf{Y}_\tau) \nonumber \\
	&\approx I(\mathbf{Y}_d;\mathbf{X}_d| \mathbf{H})  \nonumber \\
	&\geq I(\mathbf{Y}_d;\mathbf{X}_d)
	\end{align}
	where $I(\mathbf{Y}_\tau;\mathbf{X}_d)=0$ because $\mathbf{X}_d$ is independent of $\mathbf{Y}_\tau$, and the equality holds if and only if $\mathbf{H}$ is fixed. Intuitively, the pilot signal block is able to mitigate the channel uncertainty to a certain level. Generally, one of the major advantages of the pilot-aided waveform design is that it bypasses channel matrix inversion or factorization which are needed in most of the channel estimation solutions. Therefore, the required computational complexity is much lower than the others. Despite, such an approach still suffers a fractional \gls{dof} loss which will directly affect the system capacity as we previously shown in \eqref{eq05}.
	
\subsection{Learning-based Waveform Design}  
\begin{defn}\label{defn3}
It is called learning-based waveform design when the multiuser waveform is partially or even completely designed by neural network. If the information-bearing symbol is directly embedded in the designed waveform, it is called systematic waveform. Otherwise, it is called non-systematic waveform, where no artificial design is involved; as shown in \figref{figure1}.
\end{defn} 
	
	For learning-base waveform design, the transmitter side is modeled as a group of linear layers, with each weighting matrix can be viewed as a user-specific codebook. The difference between systematic and non-systemic waveform is the number of learnable parameters (i.e. non-systematic symbols) inside the weighting matrix. These parameters will be optimized by the model during the training procedure. In systematic waveform, the total number of learnable parameters is $(L)\times (T-1)$, together with the $(L)\times (1)$ systematic part, forms the weighting matrix $\mathbf{W}_m^{(\text{s})},~_{1\leq m\leq M}$. On the other hand, the total number of learnable parameters for non-systematic waveform is $(L)\times (T)$ since it does not contain any artificial design. The input to the linear layer is again a one-hot vector $\mathbf{s}_m,~_{1\leq m\leq M}$, with the size of $(L) \times (1)$. And the output of the linear layer is again one of its column elements. By normalizing the signal power, the designed signal can be transmitted through wireless channel. Besides, the \gls{dnn}-based noncoherent receiver remains unchanged as we have introduced in the previous section. 
	
	\begin{prop}[see \cite{8437142}]\label{prop2}
		The goal of multiuser joint waveform design is to find a set $\matc{\mathcal{A}}=\left \{\mathbf{\Theta}_1,\dots,\mathbf{\Theta}_J\right \}$ (or equivalently $\mathbf{C}_m,~_{\forall m}$) that minimizes the error probability. 
	\end{prop}
	
	Given the waveform set to be equally probable, the probability of signal $\mathbf{\Theta}_i$ being detected incorrectly in the noise-free case can be expressed as
	\begin{equation}\label{eq09}
	\mathcal{P}_\epsilon = \mathbb{E}\left ( \frac{\sum_{i\neq j} p(\mathbf{\Theta}_j^{-1}\mathbf{\Theta}_i\mathbf{H}_i)}{\sum_{j=1}^{J} p(\mathbf{\Theta}_j^{-1}\mathbf{\Theta}_i\mathbf{H}_i)} \right )
	\end{equation}
	where $\mathbf{H}_i\triangleq \mathbf{\Theta}_i^{-1}\mathbf{Y}$ in the noiseless case. Thus, the codebook design aims to minimize the following objective function
	\begin{equation}\label{eq10}
	\underset{\matc{\mathcal{A}}}{\min}~\mathcal{P}_\epsilon=\underset{\matc{\mathcal{A}}}{\min}~(1-\Gamma(\mathbf{H}_i,\mathbf{\Theta}_i))
	\end{equation}
	where
	\begin{equation}\label{eq11}
	\Gamma(\mathbf{H}_i,\mathbf{\Theta}_i) \triangleq \mathbb{E}\left ( \frac{p(\mathbf{H}_i)}{\sum_{j=1}^{J} p(\mathbf{\Theta}_j^{-1}\mathbf{\Theta}_i\mathbf{H}_i)}  \right )
	\end{equation}
	In general, this optimization problem is mathematically intractable and highly depends on the probability distribution of $\mathbf{H}_i$. We might employ Cauchy\textendash Schwarz inequality to obtain
	\begin{equation}\label{eq12}
	\Gamma(\mathbf{H}_i,\mathbf{\Theta}_i) \leq \sqrt{\mathbb{E}\left(\frac{p(\mathbf{H}_i)}{(\sum_{j=1}^{J} p(\mathbf{\Theta}_j^{-1}\mathbf{\Theta}_i\mathbf{H}_i))^2}\right)\mathbb{E}\left(p(\mathbf{H}_i)\right)}
	\end{equation}
	and the upper bound can be achieved at
	\begin{equation}\label{eq13}
	p(\mathbf{H}_i) + \sum_{i\neq j}^{} p(\mathbf{\Theta}_j^{-1}\mathbf{\Theta}_i\mathbf{H}_i) = \lambda
	\end{equation}
	where $\lambda$ is a constant number. Generally, it is mathematically challenged to obtain sufficient conditions for \eqref{eq10} and the optimization problem in \eqref{eq13} is an integer linear programming (ILP) problem which is NP hard. The above motivates us to utilize deep learning technique to find an acceptable waveform set $\matc{\mathcal{A}}$. Moreover, since the proposed deep learning algorithms do not have pilot-overhead, it is trivial to justify that the achievable \gls{dof} for communication is $M$ without any fractional loss.

	\subsection{Complexity Analysis}
	The computational complexity required for training the pilot-aided waveform design approach is $\mathcal{O}(|\mathcal{B}|N^2)$ per iteration, where $|\mathcal{B}|$ is the size of mini-batch; and $\mathcal{O}(N^2)$ per detection after training finished. The complexity is dominated by the matrix multiplication since the proposed network architecture only involves a feedforward neural network with several layers. The learning-based waveform design has the same complexity of $\mathcal{O}(N^2)$ since it has similar architecture with the pilot-aided waveform design. To put this in perspective, the \gls{ls} channel estimation with \gls{zf} equalization has a complexity of $\mathcal{O}(N^3)$ as matrix inversion is needed, but it is non-iterative and no training is required. Besides, the \gls{ls} channel estimation with \gls{mlsd} has a complexity of $\mathcal{O}(L^M)$.
	
	\section{Simulation results and evaluation} 
	\subsection{Implementation Details}
	For pilot-aided waveform design, the training data consists of a number of randomly generated pairs $\mathbf{p}^{(i)}=\left \{\mathbf{s}^{(i)}, \mathbf{y}^{(i)}\right \}$, where $\mathbf{y}^{(i)}$ is obtained by transmitting pilot-aided signal blocks $\mathbf{X}^{(i)}$ through a random Rayleigh block-fading channel $\mathbf{H}^{(i)}$, and $\mathbf{s}^{(i)}$ is the referenced training target as we introduced in Section III-A. For learning-based waveform design, the training input is a group of one-hot vectors $\mathbf{s}_m^{(i)},~_{1\leq m \leq M}$, and the supervisory training output is again $\mathbf{s}^{(i)}$ by using codebook combination approaches \cite{2020arXiv200400404X}.
	
	The neural network is trained by using \gls{sgd} algorithm with Adam optimizer at early training stage. After a certain number of training epochs, we removes the Adam optimizer and only \gls{sgd} algorithm is used to train the network until converging since \gls{sgd} is believed to offer better convergence performance during final training phase \cite{2018arXiv180410587B}. The learning rate is set to be 0.001 and the size of mini-batch is 200. The loss function is categorical cross entropy which is defined as
	\begin{equation}\label{eq14}
	\mathcal{L}(\mathbf{s}^{(i)},\hat{\mathbf{s}}^{(i)})=-\frac{1}{|\mathcal{B}|}\sum_{\mathbf{p}^{(i)}\in \mathcal{B}}^{} \mathbf{s}^{(i)}\log \hat{\mathbf{s}}^{(i)} 
	\end{equation}
	where $\mathcal{B}$ is the training set of a mini-batch.
	
	\begin{figure}[!t]
		\centering
		\includegraphics[width=3.3in]{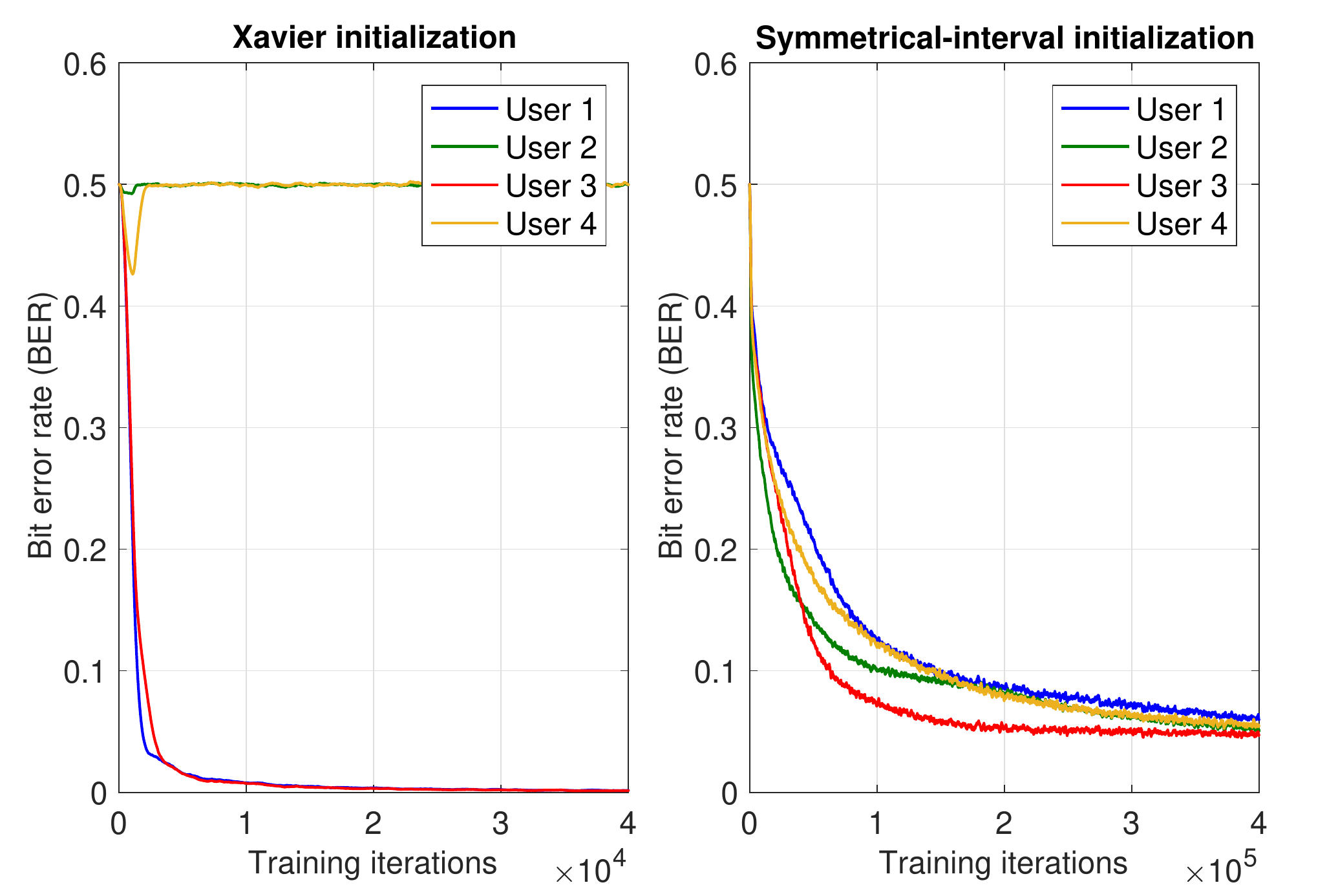}
		\caption{The convergence performance of the Xavier initialization (left) and the proposed symmetrical-interval initialization method (right).}
		\label{figure2}
	\end{figure}
	
	In addition to these basic settings, here we also propose a novel weight initialization method for the learning-based waveform design. It is well known that weight initialization plays an important role in contributing to the \gls{dnn} convergence performance. The traditional method, such as Xavier \cite{pmlr-v9-glorot10a}, is to generate weights by using the following heuristic
	\begin{equation}\label{eq15}
	\mathbf{W} \sim U \left ( 0,\frac{1}{\sqrt{n}} \right ) 
	\end{equation}
	where $U \left ( a,b \right )$ denotes the uniform distribution in the interval $\left( a-b,a+b \right )$ and $n$ is the size of the input. However, such a method could result in a user imbalance problem. In \figref{figure2} (left), we found that two users fails to converge as their detection error rate remains unchanged at 0.5. This is because the initialized weighting matrix $\mathbf{W}_m,~_{1\leq m\leq M}$ might have its column elements differ by orders of magnitude before entering the normalization function. Therefore, individual user's waveform dominates the entire transmitted signal in training phase. To tackle this issue, we proposed a novel weight initialization method, termed symmetrical-interval initialization which randomly generates the weighting matrix elements within a pair of symmetrical intervals as described in the following equation
	\begin{equation}\label{eq16}
	\mathbf{W} \sim U \left ( \Big( -\frac{1}{\sqrt{n}} ,\zeta \Big )\cup\Big ( \frac{1}{\sqrt{n}} ,\zeta \Big ) \right ) 
	\end{equation}
	
	where $\zeta$ is an arbitrary constant one order of magnitude smaller than $1/\sqrt{n}$. By such means, the convergence performance of the proposed learning-based waveform design has been greatly improved as shown in \figref{figure2} (right).
	\begin{table}[!t]
		\footnotesize
		\centering
		\caption{Layout of the \gls{ann}}\label{table1}
		\renewcommand{\arraystretch}{1.2}
		\begin{tabular}{p{3cm}|p{3cm}}
			\hline
			\textbf{Layer} &\textbf{Output dimension} \\ \hline
			Input  & $2NT$ \\ 
			Dense + ReLU & $1024$   \\ 
			Dense + ReLU & $512$   \\
			Dense + Sigmoid & $M\log L$   \\ \hline
		\end{tabular}
	\end{table}
	
	All simulations are run on a Dell PowerEdge R730 2x 8-Core E5-2667v4 server, and implemented in Matlab.
	\subsection{Simulation and Performance Evaluation}
	Our computer simulations are structured into two parts with respect to the size of \gls{mumimo} network. For pilot-aided waveform design, the number of pilot symbols sent by each user is equal to the user number $M$ in the simulation (i.e. $T_\tau = M$). For learning-based waveform design, the coherent interval $T$ is set to be the same as the previous case to ensure the fair performance comparison. The modulation scheme is set to be the quadrature phase-shift keying (QPSK). The key metric utilized for performance evaluation is the average \gls{ber} over sufficient Monte-Carlo trails of block Rayleigh fading channels. The \gls{snr} is defined as the average received information bit-energy to noise ratio per receive antenna for the entire $T$ coherent intervals. Besides, the neural network layout is listed in Table I.
	
	\begin{figure}[!t]
		\centering
		\includegraphics[width=3.3in]{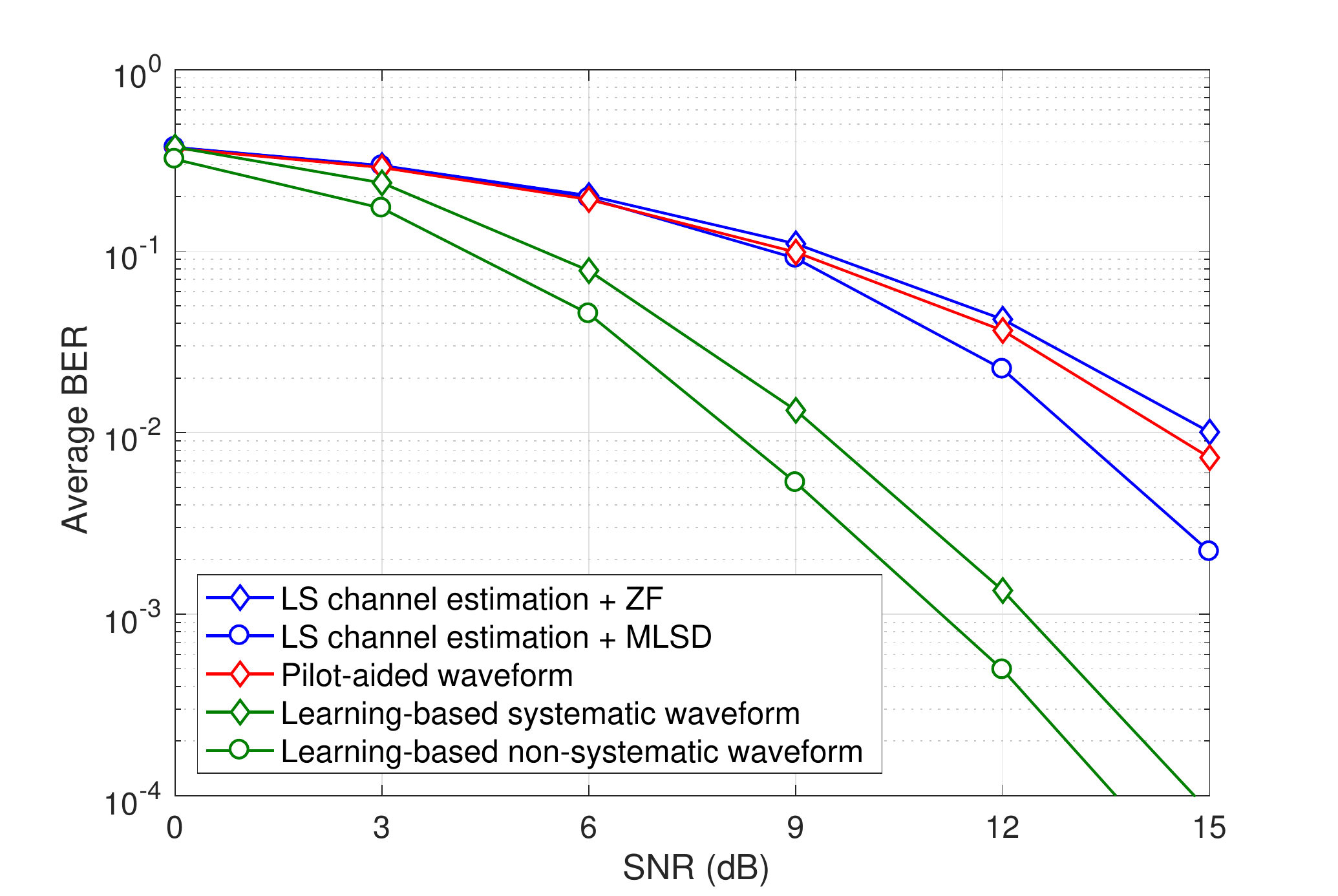}
		\caption{Performance comparison among various schemes with $M=4$, $N=8$, $T_\tau=4$, $T_d=1$.}
		\label{figure3}
	\end{figure}

	\begin{figure}[!t]
		\centering
		\includegraphics[width=3.3in]{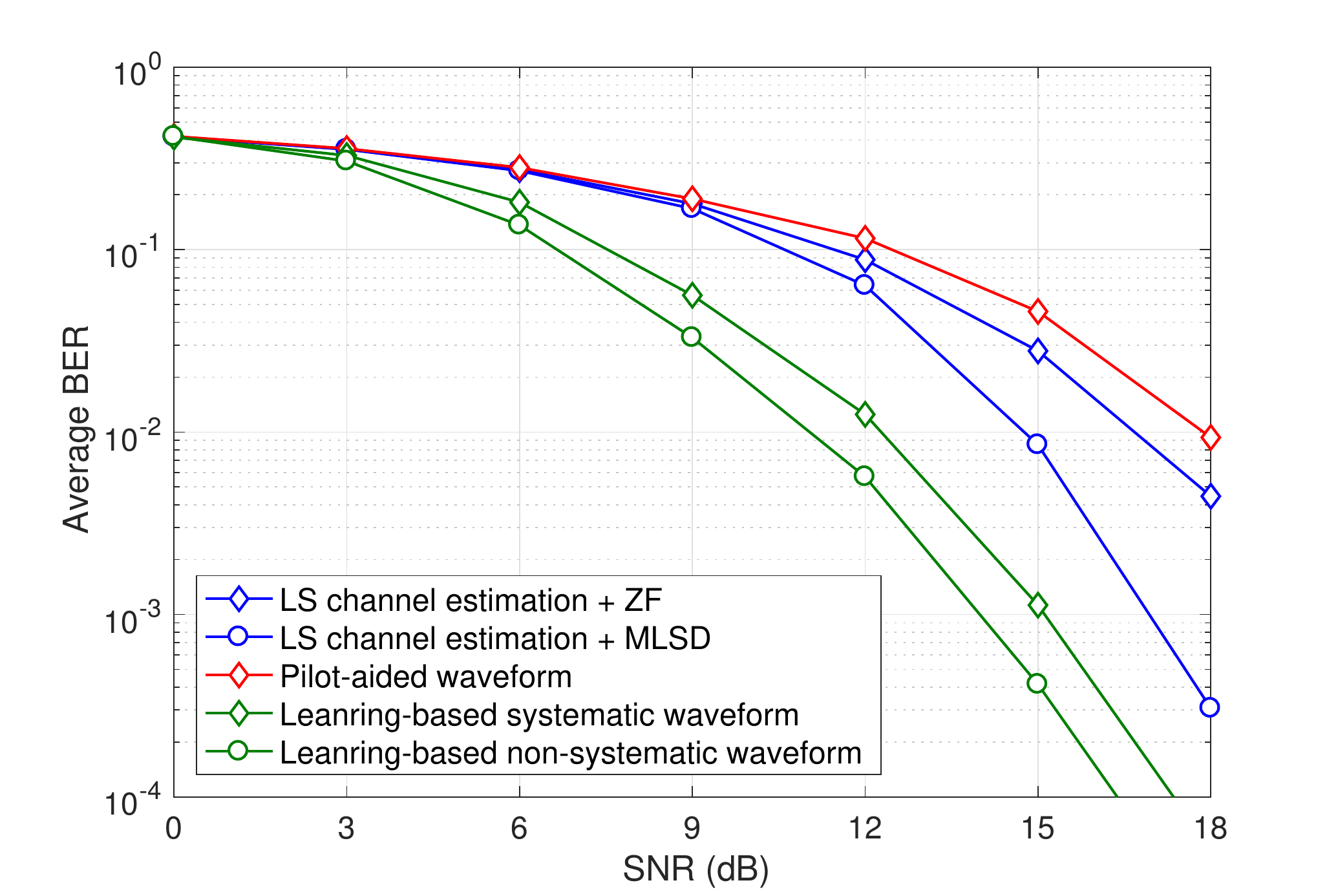}
		\caption{Performance comparison among various schemes with $M=8$, $N=16$, $T_\tau=8$, $T_d=1$.}
		\label{figure4}
	\end{figure}
	
	\figref{figure3} shows the average \gls{ber} performance of the proposed deep learning approaches in 4-by-8 \gls{mumimo} system. The baselines for performance comparison are obtained by using the conventional \gls{ls} channel estimation with \gls{zf} equalization or \gls{mlsd}. The coherent time slot $T$ is set to be 5. It is shown that the proposed pilot-aided waveform design slightly outperforms the conventional \gls{ls} channel estimation with \gls{zf} equalization. The performance improvement is around 0.6 dB at \gls{ber} of $10^{-2}$. But the gap to the \gls{ls}-\gls{mlsd} solution is still around 1.5 dB at \gls{ber} of $10^{-2}$. Meanwhile, two learning-based waveform design approaches demonstrate their tremendous advantages towards the conventional pilot-based algorithms. The \gls{ber} improves about 3.7 dB for systematic waveform and 5 dB for non-systematic waveform at \gls{ber} of $10^{-3}$, respectively. This is reasonable since the time-domain redundancy obtained by removing the pilot symbols has been well exploited. Moreover, the non-systematic approach outperforms the systematic approach for approximately 1 dB at \gls{ber} of $10^{-3}$ since the former provides higher flexibility for joint multiuser codebook design.
	
	\figref{figure4} shows the average \gls{ber} performance of the proposed deep learning approaches in 8-by-16 \gls{mumimo} system. The baselines for performance comparison remain unchanged. The coherent time slot $T$ is set to be 9. Different from the previous case, here the performance of the pilot-aided waveform slightly decreases with a \gls{ber} gap around 1.3 dB compares with the conventional \gls{ls}-\gls{zf} solution. This is due to the increasing \gls{iui}, which makes sequence-level signal detection more difficult. Meanwhile, the proposed learning-based approaches maintain a good performance which demonstrates their scalability with respect to the size of the \gls{mumimo} network. The \gls{ber} gap between systematic waveform and non-systematic waveform to the conventional \gls{ls}-\gls{mlsd} solution is approximately 3 dB and 2 dB at \gls{ber} of $10^{-3}$, respectively.
	
	\section{Conclusion}
	This paper aims to handle the joint transmitter and noncoherent receiver design for \gls{mumimo} systems.
	It has been shown that \gls{dnn} receiver can realize noncoherent signal detection directly under the conventional pilot-based signal model. Without the requirement of channel inversion, it can outperform the \gls{ls} channel estimation with \gls{zf} equalization in small-size \gls{mumimo} systems. The performance can be further improved by adopting the learning-based waveform design, where the transmitted waveforms are jointly optimized by deep learning algorithms. By such means, the detection performance can be further improved for more than 5 dB at high \gls{snr} range in various communication scenarios.

	{
		\section*{Acknowledgement}
		The work was supported in part by European Commission under the framework of the Horizon2020 5G-Drive project, and in part by 5G Innovation Centre (5GIC) HEFEC grant.
	}

	\ifCLASSOPTIONcaptionsoff
	\newpage
	\fi
	
	\balance
	\bibliographystyle{IEEEtran}
	\bibliography{./ref}
\end{document}